# Misfit layer superconductors, tuneable bulk heterostructures with strong 2D effects

T Samuely[1], M. Gmitra[2], T Cren[3], M Calandra[4], P Samuely[2]

[1]P.J. Šafárik University, 04001 Košice, Slovakia
[2]Institute of Experimental Physics, Slovak Academy of Sciences, 04001 Košice, Slovakia
[3]University of Trento, 38123 Povo, Italy
[4] Sorbonne Université, CNRS, Institute des Nanoscience de Paris, F-75252 Paris, France

Atomically thin layered materials are systems with zero limit bulk-to-surface ratio. Their physical properties are determined by two-dimensionality (2D) and strongly affected by interfacing with other systems. Therefore, they represent an accessible platform for the abundance of quantum effects that can be engineered by combining them into vertical stacks. Two types of layered systems are considered here – artificially prepared (exfoliated) van der Waals nanostructures [1], and naturally layered systems showing quasi 2D behaviour already in a bulk form. A special class of naturally layered materials is misfit structures combining atomic layers of hexagonal (H) transition metal dichalcogenides (TMD) and slabs of tetragonal (T) ionic rare-earth monochalcogenides in the same superlattice [2]. Both types of layered systems feature a new state of quantum matter, the Ising superconductivity extremely resilient to external magnetic field. A giant electron doping, natural to the misfit structures, can lead to topological superconductivity. Both systems can also be assembled into heterostructures combining different constituents. Layered 2D heterostructures have a large number of implications for many potential applications in solid-state devices, solar cells, photodetectors, semiconductor lasers, light-emitting diodes, and biosensors [3]. One of the most challenging applications of topological superconductors is quantum computation with putative Majorana zero-energy modes which are quasiparticles with the properties of non-Abelian anyons. Braiding of such quasiparticles constitutes the basis of topologically protected qubits, a robust solution to the problem of decoherence and unitary errors of the state-of-the-art quantum computers. In spite of extensive theoretical elaboration, the physical realisation of the topological quantum bits is in its infancy, and fundamental research in the field of topological superconductivity is needed. This includes also understanding of physical properties of the materials used as building blocks of artificial heterostructures.

**Current and Future Challenges**
One way to topological superconductivity is Ising superconductivity (IS). IS has been discovered in atomically thin TMD films, namely in ion-gated $MoS_2$ and in monolayer $NbSe_2$ [4] showing extremely large in-plane upper critical magnetic fields much above the theoretical Pauli limit, where the Zeeman energy equals to condensation energy. Within a single TMD layer, finite in-plane electric fields are allowed by broken inversion symmetry leading through strong spin-orbit interaction to effective *k*-dependent magnetic fields polarized perpendicularly to the layer locking the orientation of the electron spin to the out-of-plane direction (Ising) and making the action of experimentally achievable in-plane external magnetic field highly irrelevant. A careful transport study of a sequence of a monolayer, double layer, etc., of $NbSe_2$ shows a suppression of the upper critical field with increasing number of layers so the effect is restricted to 2 dimensions.

Our team studied the family of bulk misfit layer compounds where mutually misfitting hexagonal $NbSe_2$ and tetragonal LaSe layers are alternating which exhibit large superconducting anisotropy and quasi 2D character. The most surprisingly even more extreme in-plane upper critical magnetic fields $B_{c2//ab}$ strongly violating the Pauli paramagnetic limit have been observed compared to monolayers. Namely, the misfit layer $(LaSe)_{1.14}(NbSe_2)$/1T1H single crystal (Fig. 1(a)) with $T_c$ = 1.23 K shows $Bc_{2//ab}$ = 20 T and the crystal $(LaSe)_{1.14}(NbSe_2)_2$/1T2H (Fig. 1(b)) with $T_c$ = 5.7 K exhibits $B_{c2//ab}$ = 50 T, which is 10 and 5 times more than the respective Pauli limit [5]. Complementary experimental methods of transport measurements, ARPES, STM and QPI in combination with first-principles calculations bring surprising evidence that it is due to IS surviving in bulk $NbSe_2$-based materials despite the fact that the in-plane inversion symmetry should be preserved [6]. It was shown that the band structure of 1T2H and 1T1H crystal is very similar to that of the monolayer $NbSe_2$ with a rigid band shift (Fig. 2). Namely, in 1T2H the LaSe layer donates about 0.6 electron per $NbSe_2$ chemical unit corresponding to the Fermi level shift of 0.3 eV in comparison with monolayer $NbSe_2$. 1T1H is even more heavily doped and the Fermi level is shifted by 0.5 eV still not completely filling the hole Nb band. Moreover, ARPES proved that the misfits are quasi-2D systems with a very little dispersion of the electronic band along the $z$-direction. Being the electronic equivalent with $NbSe_2$ monolayer with spin-split bands around the K(K') points in the Brillouin zone makes the misfits Ising superconductors. In [7] a theoretical model shows that in 1T1H the coupling between the superconducting $NbSe_2$ layers through the insulating LaSe is so small that 1T1H creates an infinite stack of monolayers, each without inversion symmetry. In 1T2H, the structure can be viewed as a stack of seemingly centrosymmetric 2H-$NbSe_2$ bilayers separated by LaSe misfit layers. But LaSe layers below and above the $NbSe_2$ bilayer are oriented differently with respect to $NbSe_2$, thus breaking the total inversion symmetry in the misfit. Hence, the hierarchy of energy scales of interlayer coupling, namely, breaking of inversion symmetry by the LaSe layer, spin-orbit coupling and the magnitude of the superconducting gap in the 1T1H and 1T2H compounds, provides a consistent picture for the strong Ising protection observed in these bulk compounds. This is different from the IS in the monolayer where only the last two energy scales are present.

**Advances in Science and Technology to Meet Challenges**
Tunability of properties is of primary importance in quantum design. The success of 2D materials in different fields of science is in their better tunability in comparison with the bulk materials. One of the knobs in doping. By an double-layer FET geometry $10^{14}$ electrons per $cm^2$ can be introduced to the monolayer of $NbSe_2$. Somewhat larger doping can be achieved via deposition of alkali atoms (such as potassium). The large electron charge transfer in the misfits allows to obtain doping fractions that largely encompass both those obtained in the case of ionic-liquid-based FETs and K adatom deposition. This large transfer from LaSe to the TMD is due to the chemical properties of the highest occupied states of bulk LaSe that are formed only by atomic La states. As a consequence, LaSe essentially behaves as a donor, an effect not depending on the kind of TMD used to build the misfit structure. This suggests that similarly large doping can be achieved by sandwiching LaSe with other metallic TMDs. It is also possible to vary the rocksalt layer, opening a tremendous space of possibilities and a fertile ground for the discovery of highly innovative materials. For instance, assuming La3+, Pb2+, and Se2− in the compounds $(La_{1−y}Pb_ySe)_{1+x}(NbSe_2)_2$, a simple charge balance calculation shows that the charge transfer from the rock salt layer $(La_{1−y}Pb_ySe)$ to the $(NbSe_2)_2$ double layer could vary continuously from 0 to 1 + $x$ by reducing the lead content. Therefore, the charge transfer could be tuned by appropriate substitution in misfit layer compounds, pretty much in the same way as doping can be varied in FET, but in a much broader range [6].

Tuneable charge doping in the misfits introduces the possibility of shifting the Fermi level to be in between the two spin-split bands, which is predicted to give rise to topological superconductivity [8]. It is plausible that these insights can be extended to other misfit compounds and bulk structures comprising TMD layers, where large in-plane critical magnetic fields that exceed the Pauli limit have been reported but not identified as evidence of Ising superconductivity. This includes works on $(SnSe)_{1.16}(NbSe_2)_2$, $Ba_6Nb_{11}S_{28}$, cation-intercalated $NbSe_2$, and organometallic intercalated compounds of 2H-$TaS_2$ (references in [7]). Future experiments and theory work to determine the electronic and structural properties for this broad range of potential bulk Ising superconductors will be of great interest.

Few layer transition-metal dichalcogenides and their misfit structures can be easily incorporated in 2D heterostructures (Fig. 1(c)). Proximity-induced spin-orbit coupling and exchange interactions in heterostructures made of 2D materials are sources of novel quantum effects engineered through elaborate stacking. Using this approach, there have been created artificial systems where proximity effects are combined to yield entirely new physical qualities (*p*-wave superconductivity in ferromagnetic/*s*-wave superconductor systems, Yu-Shiba-Rusinov states, Majorana bound states, etc.) [9].

**Concluding Remarks**
Abundance of highly innovative materials of untapped properties can be created in the form of misfit layer compounds. Their unit cells are naturally growing heterostructures of different stacking from tetragonal rocksalt layers and hexagonal TMD with various electronic and magnetic properties. Unpredicent tuning of their properties is controlled by the charge transfer/doping from T layers to H layers. Concerted effect of charge-transfer, defects, reduction of interlayer hopping, and stacking enables Ising superconductivity. It provides a possible pathway to design of bulk superconductors that are resilient to magnetic fields. Ising spin-orbit coupling can also be tuned to topological superconductivity. Like traditional layered materials, the misfits are often exfoliatable and incorporatable as units of artificially stacked heterostructures.

**Acknowledgements**
This work was supported by Projects COST Action No. CA21144 (SUPERQUMAP), APVV-20–0425, APVV-SK-FR-22-0006, VEGA 2/0073/24, Slovak Academy of Sciences IMPULZ IM-2021–42.

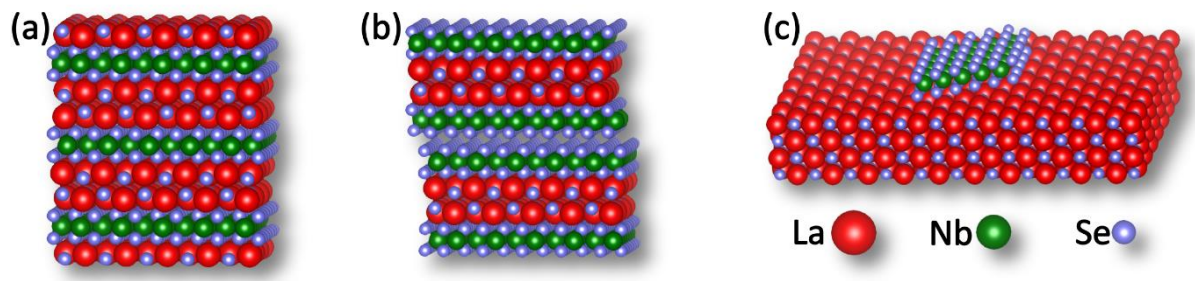

**Figure 1** Crystal structure of misfit layer superconductors. (a) 1T1H or (LaSe)$_{1.14}$(NbSe$_2$) ionocovalent crystal with $T_c$ = 1.3 K and in-plane upper critical field of 20 T. (b) 1T2H or (LaSe)$_{1.14}$(NbSe$_2$)$_2$ where HTH trilayers are van der Waals bound, with $T_c$ = 5.7 K and in-plane upper critical field of 50 T. (c) Artificial heterostructure where a monolayer of hexagonal NbSe$_2$ is placed on top of tetragonal LaSe crystal.

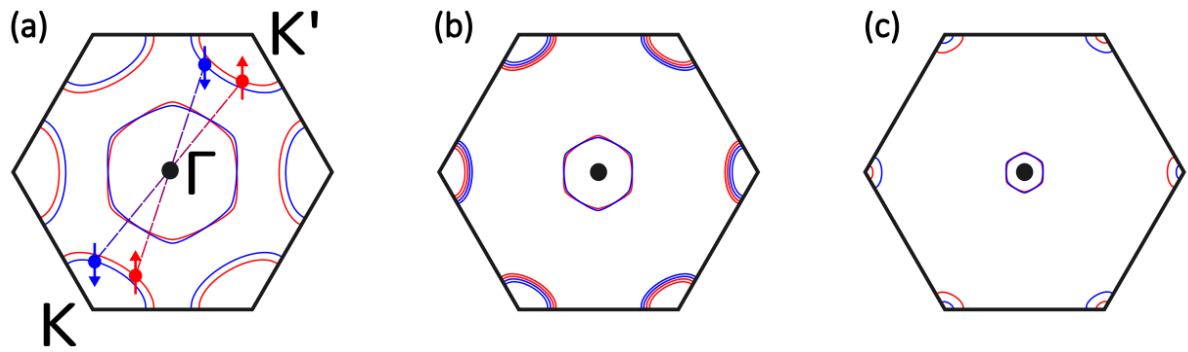

**Figure 2** Brillouin zone and Fermi surfaces of (a) monolayer $NbSe_2$, (b) 1T2H or $(LaSe)_{1.14}(NbSe_2)_2$ and (c) 1T1H or $(LaSe)_{1.14}(NbSe_2)$ showing gradual shrinking of the hole Fermi surfaces due to doping of $NbSe_2$ layers from LaSe. Spin splitting around K and K' points is responsible for Ising superconductivity.